\documentclass{aa}  
\usepackage{graphicx}
\usepackage{physics,amsmath}
\usepackage{txfonts}
\usepackage{placeins}
\usepackage{xcolor}
\usepackage{float}

\usepackage{natbib}
\bibpunct{(}{)}{;}{a}{}{,} 
\usepackage[colorlinks=true, allcolors = blue]{hyperref}

\hypersetup{
  colorlinks   = true,
  urlcolor     = blue,
  linkcolor    = blue,
  citecolor   = blue 
}
\usepackage{xcolor}

\begin{document} 

   \title{Simulations of massive star atmospheres and winds during giant eruptive and quiescent luminous blue variable phases}
   \author{P. Schillemans, J.O. Sundqvist, D. Debnath, L. Delbroek, N. Moens, C. Van der Sijpt}

   \institute{Institute of Astronomy, KU Leuven, Celestijnenlaan 200D, 3001, Leuven, Belgium,\\
    \email{pieter.schillemans@kuleuven.be}}

   \date{Received 23 March 2026; Accepted XXX}

   \titlerunning{Simulations of massive star atmospheres and winds during giant eruptive and quiescent LBV phases}
   \authorrunning{P. Schillemans et al.}

  \abstract
   {Mass loss from massive stars located in the part of the Hertzsprung-Russell diagram (HRD) where we find luminous blue variables (LBVs) is profoundly important for stellar evolution, yet very poorly understood.} 
   {We use time-dependent radiation-hydrodynamic (RHD) simulations to examine the atmosphere and wind properties of massive stars in this region of the HRD.} 
   {We compute 2D and 1D RHD models of the coupled envelopes, atmospheres, and wind outflows of massive stars, tuned to the region in the HRD where LBVs are located. Our unified simulations, which start deep in the stellar envelope (below the iron opacity bump) and range well into the outflowing wind, account for line-driving as well as radiative enthalpy and photon tiring. Mass-loss rates and wind speeds are thereby emergent properties in the simulations. A grid of models is created by slightly increasing the stellar energy at the lower boundary, which then controls the emergent radiative luminosity at the photosphere.} 
   {Increasing total stellar energy results in a natural transition from very turbulent atmospheres with line-driven winds to effectively stationary super-Eddington massive outflows. In the latter case, radiative enthalpy keeps the star below the Eddington limit in the deep layers where the wind is launched. However, as energy transport by enthalpy becomes inefficient as we approach the surface, the star is observed to be highly super-Eddington at its recombined photosphere. Our sub-Eddington models are essentially blue hypergiant stars ($T_{\rm eff} \sim 10~000-20~000$ K) with very variable surfaces, effective mass-loss rates $\dot{M} \sim 2-5 \times 10^{-5} \ \rm M_\odot/year$, and wind speeds $\varv_\infty \sim 200-300$ km/s, resembling quiescent LBVs like P Cygni. The super-Eddington models have optically thick wind envelopes and extremely inflated yellow surfaces ($T_{\rm eff} \sim 5000$ K), $\dot{M} \sim 0.1-1 \ \rm M_\odot/year$, and $\varv_\infty \sim 400-500$ km/s, resembling a massive star during a great eruption such as that of $\eta$ Carinae. } 
   {Although not tuned to any specific star, our RHD atmosphere and wind models naturally reproduce the overall characteristic stellar and wind parameters inferred for massive stars in their quiescent LBV and yellow giant eruptive phases. It remains an open question whether the energy increase needed to trigger a giant eruption can be obtained solely by the internal evolution of the star itself or if it requires an external energy source such as, e.g., a stellar merger or a close binary interaction.}


   \maketitle

\section{Introduction}

As massive stars evolve, they tend to do so toward lower effective temperatures and higher luminosities, while losing a significant amount of mass in radiation-driven winds. This means that they evolve upwards and to the right in the Hertzsprung-Russell diagram (HRD), and go closer and closer to the classical Eddington limit.
As a result, very massive stars can undergo very unstable and even eruptive phases in their lifetimes (e.g. \citealt{humphreys_davidson_94,owocki_gayley_97, smith_owocki_2006,smith_2014}). 
For example, S Doradus-type luminous blue variables (LBVs) undergo semi-periodic brightness changes, which are also connected to changes to the observable mass-loss rate. These can vary up to two orders of magnitude, over the course of years or decades (e.g. \citealt{groh_2009,grafener_2012, grassitelli_2021}). Other classes, such as the (poorly defined) supernova impostors and giant eruption objects, like the famous $\eta$ Carinae eruption, experience violent and sudden eruptive events. During such episodes, their brightness increases 4 -- 10 mag, and up to several solar masses are lost in singular events lasting months to decades (e.g. \citealt{smith_2011, rest_2012}). 
This same HRD regime is additionally relevant for the early Universe, as both theoretical models and recent observations with the James Webb Space Telescope (e.g. of so-called `Little Red Dots') hint at the presence of extremely massive Population III stars  (e.g. \citealt{schaerer_2002, takashi_2013,santarelli_2026,nandal_2026}). These supermassive first stars
are predicted to live in the same corner of the HRD, resulting in, at the very least, massively inflated atmospheres.

The specific physical processes governing the coupled stellar envelope, atmosphere, and wind outflow in this regime are still poorly understood. It is, for example, unclear to which extent the transition from a semi-steady state to S Doradus-type stars or to super-Eddington eruptive events can occur naturally in (single-star) evolution (e.g. \citealt{langer_1994,meynet_maeder_2005,aadland_2018}, but also \citealt{groh_2013,smith_2015}), and what physical processes then are responsible, or if additional energy input is always required, e.g. from mergers or other binary interactions (e.g. \citealt{justham_2014,aghakhanloo_2017,Hirai21,mahy_2022}).

Additionally, even classifying these stars is challenging, as confirmation of objects as LBVs requires the observation of either S Doradus-type variability or a giant eruption, since they do not exhibit unique spectroscopic or photometric diagnostics \citep[e.g.][]{humphreys_davidson_94,weis_bomans_2020}. This makes identification and definitive classification extremely difficult, with only some 40 LBVs and some 100 LBV candidates known. 

Radiative-hydrodynamical (RHD) simulation efforts that are able to reproduce the properties of such unstable and eruptive stages in this region of the HRD are largely lacking. Previous multi-dimensional RHD simulations typically have been tuned toward lower luminosities or other massive-star evolution stages \citep{jiang_2015,jiang_2018,nico_2022b,nico_2025,Debnath24,goldberg_2022,goldberg_2025}, whereas 1D super-Eddington models often have neglected important physical processes. In this latter category we find \citet{Quataert2016, Owocki2019}, where super-Eddington winds were modelled in an adiabatic way, ignoring the effects of the radiative flux on the resulting structure, and \citet{owocki_2017}, who include both diffusive and advective flux, but whose stationary models remain parametrised in the sense that they neglect the important effect of hydrogen recombination by assuming a constant electron scattering opacity. 
    
Modelling this parameter space is generally very important in the evolution of very massive stars, as the effects of e.g. mass-loss rates, sudden eruptive stages, etc, on their end-of-life products are huge \citep{smith_2014}. In this paper, we describe a series of 2D and 1D RHD models with increasing stellar energies in this challenging corner of the HRD that include gas and radiative enthalpy, co-moving radiative flux, and line-driving effects, building on the techniques described in detail in \citet{luka_2022,nico_2022a,nico_2022b,Debnath24}. The series of models shows a clear transition with the increase in stellar energy, from rather `well-behaved' (i.e. non-eruptive) but strongly variable blue hypergiant stars, to super-Eddington objects with extreme mass loss and yellowish surfaces, pointing to a natural transition between the variety of objects found in this corner of the HRD. In Sect. \ref{Sect:Method}, we briefly go over the modelling specifics including the initial conditions, as well as provide a description of the grid of models. In Sect. \ref{Sect:Results}, we show and analyse two representative models, one before and one after the super-Eddington transition, and examine their differences. In Sect. \ref{Sect:Discussion}, we further discuss the results, give our conclusions and an outlook for future work.

\section{Method} \label{Sect:Method}

Building on the work done by \citet{nico_2022b,Debnath24}, we solve the time-dependent RHD equations using the RHD module \citep{nico_2022a} in \texttt{MPI-AMRVAC} \citep{xia_2018,Keppens_2023}. These equations, namely the conservation of mass, momentum and energy of the gas, are solved on a finite volume grid in a `box-in-a-star' approach, including spherical divergence correction terms (see Appendix A in \citealt{nico_2022a}). The radiation field is included in both the momentum equation, as a radiative force, and in the energy equation, as radiative heating and cooling. The code properly treats the effects of radiative enthalpy and so-called `photon tiring' \citep{owocki_gayley_97}. The radiation properties (i.e. the radiation energy density, flux, and pressure, as well as the coupling with the gas) are calculated by solving the frequency-integrated zeroth moment of the time-dependent radiative transfer equation in the co-moving frame (CMF). The system of gas equations is closed by the ideal gas law assuming a constant gas adiabatic index $\gamma_g = 5/3$ (see also \citealt{jiang_2015, goldberg_2022}). 

For the radiation field, we use the flux-limited diffusion (FLD) approximation as outlined in \citet{nico_2022a} as our closure relation. This closure is analytic in the optically thick (diffusive) and optically thin (free-streaming) limits, and uses an approximate bridging law in between. For the fully written-out equations, we refer to Eqs. 1 -- 9 in \citet{Debnath24}. As in preceding work by \citet{nico_2022a,nico_2022b,Debnath24}, the energy-, flux-, and Planck-weighted mean opacities are all taken to be equal to a total opacity $\kappa$. This total opacity is the so-called `hybrid opacity', as introduced by \citet{luka_2022}, which includes both Rosseland mean opacities, here in the form of the tabulated OPAL opacities \citep{opal} complemented with low-temperature opacities from \citet{ferguson_2005}, and line opacities computed in the \citet{sobolev_1960} approximation that account for line-driving \citep{cak1975, gayley_1995,luka_2022}. Included in the treatment of the line force is the finite-disk correction factor \citep{pauldrach_1986,friend_1986}, as implemented in \citet{Debnath24}.
In contrast to \citet{nico_2022b, Debnath24}, in this paper we opted to keep the line-force parameters fixed (to $\bar{Q} = Q_0 = 2000$ and $\alpha = 2/3$) instead of letting them vary according to the local atmospheric conditions. 
This produces more stable line-driving conditions in this challenging stellar regime; we discuss potential impacts of this approximation in Sect. \ref{Sect:Discussion:Limitations}.

\subsection{Initial conditions}

\begin{figure}
\centering
\includegraphics[width=\hsize]{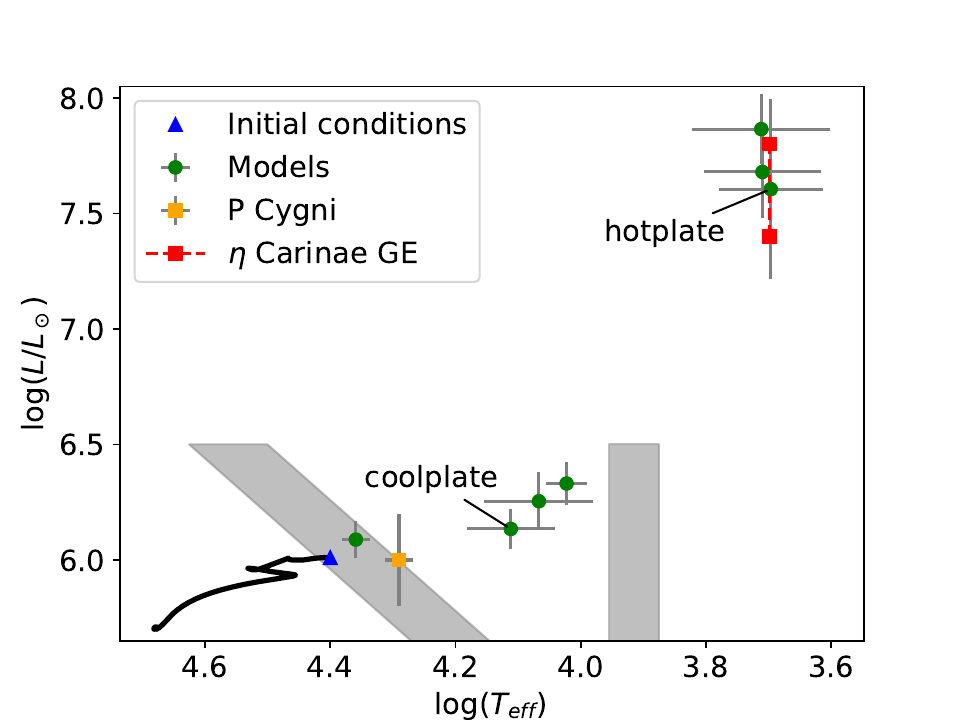}
  \caption{HRD showing the track of the \texttt{MESA}-model (black) up to the initial conditions and the resulting positions of the seven models discussed in this paper, with the two discussed in greater detail marked separately. Also shown are the approximate locations of S Doradus-type LBVs (grey), P Cygni, and the $\eta$ Carinae giant eruption, to guide the eye.}
     \label{Fig:HRD}
\end{figure}

To ensure that our initial conditions describe a physically relevant parameter space, we used one of the models of \citet{robin_2021}, who uses the stellar evolution code \texttt{MESA} \citep{Paxton2011,Paxton2013,Paxton2015,Paxton2018,Paxton2019}.
As our prototypical case, we chose a galactic 60~$M_\odot$ star, which we let evolve until it approximately reached the LBV instability strip \citep{wolf_1989}. The resulting star then had a total remaining mass of 57~$M_\odot$, its position in the HRD is shown in Fig. \ref{Fig:HRD}.

To produce initial conditions for the RHD simulations, we then took lower boundary values from the stellar structure predicted by \texttt{MESA}. Specifically, we fixed the density in the quasi-hydrostatic regime of the stellar atmosphere, well below the photosphere at a temperature of 400~kK. This way, the full 2D simulation encompasses the unstable so-called iron opacity bump region, located around a temperature of 200~kK, which is extremely important for the formation of envelope turbulence (see e.g. \citealt{Cassandra_2025}). These values, together with the radius at which this temperature and density occur, being $R_c = 38.6~R_\odot$, set the lower-boundary conditions for our time-dependent RHD simulations.
While this lower boundary remains fixed, we do not use the full structure from \texttt{MESA} as initial conditions but rather the envelope and wind conditions from our own 1D code. These initial conditions are calculated in a similar way as originally described in Sect. 2 of \citet{santolaya} (1D model atmosphere + wind code {\sc fastwind}). They assume a 1D spherically symmetric envelope in quasi-hydrostatic equilibrium, to which an analytic wind with a $\beta$-law velocity profile is connected. As in \citet{Debnath24}, we extend the atmosphere to beneath the iron bump and optionally include turbulent pressure in the momentum balance and energy transport by convection by using a standard mixing-length treatment (MLT, described in e.g. \citealt{kip_2013}) in this 1D approach. 
The advantage of calculating the initial conditions in this way is that the stellar evolution code ensures that the values are consistent with stellar structure and evolutionary constraints, while the use of a dedicated 1D atmosphere + wind code allows for better starting conditions in
the upper atmospheric and wind layers.
Specifically, the \texttt{MLT++} formalism in \texttt{MESA} artificially enhances the efficiency of convective energy transport (here by a factor of almost 30) in order to avoid density inversions and the drastic inflation of the static envelope that otherwise occurs for stars close to the classical Eddington limit \citep{grafener_2012}. 
In our code, we instead chose a standard efficiency value of $\alpha_{\rm MLT} = 1$. This means that radiation has to transport a larger fraction of the total energy, so the atmosphere in our initial conditions is inflated up to the point of the assumed launch of the stellar wind (which in our initial condition has the mass-loss rate $\dot{M} = 8.4 \times 10^{-6}~M_\odot/{\rm yr}$, stemming from the \citet{robin_2021} recipe).

\begin{figure}
\centering
\includegraphics[width=\hsize]{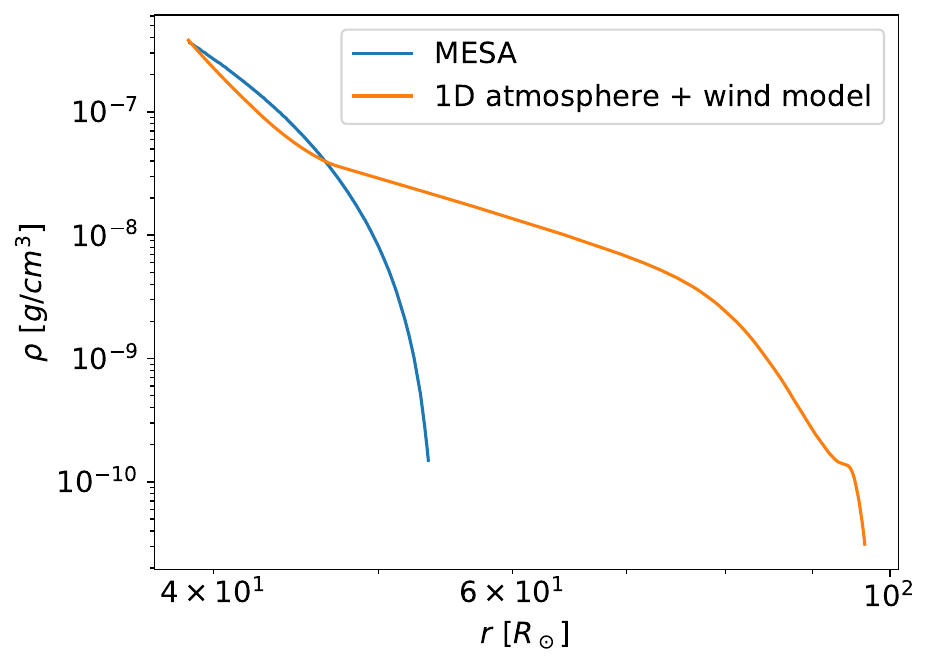}
  \caption{Comparison between the structure up to the photosphere of the 1D atmosphere + wind model and \texttt{MESA}.
  }
     \label{Fig:initial_struct_comparison}
\end{figure}

Fig. \ref{Fig:initial_struct_comparison} compares the initial structures predicted by \texttt{MESA} and our 1D envelope + analytic wind code. We can see that there is a very noticeable difference between them in this regime, with the photospheric radius increasing with nearly a factor two. 
Since the full 2D RHD simulation inflates even more during relaxation, it is clear that our choice produces initial conditions that seem to be more physical in this regime. As a bonus, this also reduces computation time needed for relaxation.

\subsection{Model grid}

To investigate a possible transition in this region from `well-behaved' stars with turbulent atmospheres to stars in outburst, we produced a grid of models with varying total stellar energy. Specifically, this was done by changing the temperature at the lower boundary $T_c$ either up or down from the initial value of 400~kK, while keeping the lower-boundary density and radius fixed. This effectively creates `hotplates' and `coolplates' for the conditions produced by the stellar evolutionary model, comparable to the methodology in the 1D work by \citet{Quataert2016,owocki_2017,Owocki2019} and the multi-dimensional red supergiant simulations by \citet{goldberg_2022}.
The model grid consists of seven models with lower-boundary temperatures that are reduced by up to 10\%, or increased by up to 15\% from the initial value of 400 kK.
This is aimed to mimic approximately the track a massive star may take on the HRD, as is visualised in Fig. \ref{Fig:HRD}. 
In this paper, we do not consider the origin of the increase in energy. An analysis whether it may arise internally from the natural evolution of a (single) star, or whether external sources such as mergers (see e.g. \citealp{owocki_2017,Owocki2019,Hirai21}) are needed, is beyond the scope of the simulations presented here. 

Importantly, our method allows us to consider the same basic star in all models, with the main alteration to the result being a different (emergent) classical Eddington factor 
\begin{equation}
\Gamma_e = \frac{g_{\rm e,c}}{g} = \frac{L_{\rm cmf}}{L_{\rm Edd}} = \frac{\kappa_{e,c} L_{\rm cmf}}{ 4 \pi c G M},
\label{Eq:Ledd}
\end{equation}
with $g_{\rm e,c}$ the radiative acceleration stemming from electron scattering, $g$ the gravitational acceleration, $\kappa_{e,c}$ the Thomson scattering opacity for a fully ionised plasma, $L_{\rm cmf}$ the CMF radiative luminosity, and $M$ the stellar mass. Equation \ref{Eq:Ledd} also defines the Eddington luminosity $L_{\rm Edd}$. As some of our models experience hydrogen recombination in their upper layers, it is important to note that $\Gamma_e$ (or equivalently $L_{\rm Edd}$) is defined only for a constant electron scattering opacity in a fully ionised plasma, $\kappa_{e,c} \approx 0.2(1+X)$ with hydrogen mass fraction $X$; in this paper we use $\kappa_{e,c} = 0.34 \rm \ cm^2g^{-1}$. 
As we will see in Sect. \ref{Sect:1D_Super-Edd_sims}, the real $\kappa_e \ll \kappa_{e,c}$ in the outer parts of our models after hydrogen has recombined, due to the lack of free electrons. We note further that the CMF luminosity is potentially different to the observer's frame luminosity $L_\ast$, 

\begin{equation}
L_\ast = L_{\rm cmf} + 4 \pi r^2 \varv(E + P),
\end{equation}

for radiation energy density $E$ and radiation pressure $P$ \citep{Cassinelli73}; the latter part essentially represents radiative enthalpy (see also \citealt{owocki_2017}).

\subsection{Simulation set-up}

As mentioned above, the 2D simulation is a so-called `box-in-a-star', using a Cartesian grid including spherical divergence correction terms, and starting from below the photosphere and extending out into the wind in one unified simulation.
The grid has periodic lateral boundaries and a fixed lower boundary at a radius $R_c = 38.6~R_\odot$. The simulations were run on a 2D grid spanning 1~$R_c$ laterally and 39~$R_c$ radially (meaning the box was placed between 1~$R_c$ and 40~$R_c$, going out to approximately 1550~$R_\odot$). We also re-run the hotplate models with $T_c \ge 440$~kK in 1D, extending the computational domain by a factor of 20. This 1D grid thus spans 799~$R_c$, going out to nearly 31~000~$R_\odot$. The reason for this, and the validity for this approximation, will be discussed in Sect. \ref{Sect:2D_Super-Edd_sims}.  

At its lowest resolution setting, every 2D simulation has a grid with 64 lateral and 1024 radial points, corresponding to cells that stretch 0.016~$R_c$ (0.60~$R_\odot$) laterally and  0.038~$R_c$ (1.47~$R_\odot$) radially. Although adaptive mesh refinement is available in \texttt{MPI-AMRVAC}, we used fixed-radius refinement, as the highly turbulent nature of the simulations would otherwise cause mesh refinement to occur over much of the domain. We used four levels of refinement, with each level doubling the grid points in both directions, at 2, 3, and 4~$R_c$. At the highest level, situated deepest in the atmosphere, there are thus 512 grid points in the lateral and 8192 in the radial direction, corresponding to cells covering 0.0020~$R_c$ (0.075~$R_\odot$) laterally and 0.0048~$R_c$ (0.184 $R_\odot$) radially. To be as close a match as possible to the 2D simulations, the 1D ones have the same resolution. So, at the lowest refinement level, 20~480 radial points are used, which is doubled three times to 163~840 at the highest setting.

The simulations presented below span nearly 1.5 years for the 2D models discussed in Sect. \ref{Sect:Sub-Edd_sims} and \ref{Sect:2D_Super-Edd_sims}, and slightly over 24 years for the 1D model discussed in Sect. \ref{Sect:1D_Super-Edd_sims}. These simulations are representative for all simulations in this respect.

\section{Results} \label{Sect:Results}

We produced in total six 2D models and additionally several 1D models in the super-Eddington regime (see above). We analyse two representative cases in more detail, one 2D model with $T_c = 380$~kK as a typical sub-Eddington turbulent model, from now on referenced as the coolplate model, and both one 2D and one 1D model with $T_c = 440$~kK as representative models with super-Eddington luminosity at the photosphere, referenced as hotplate models.

\begin{figure*}
\centering
\includegraphics[width=\textwidth]{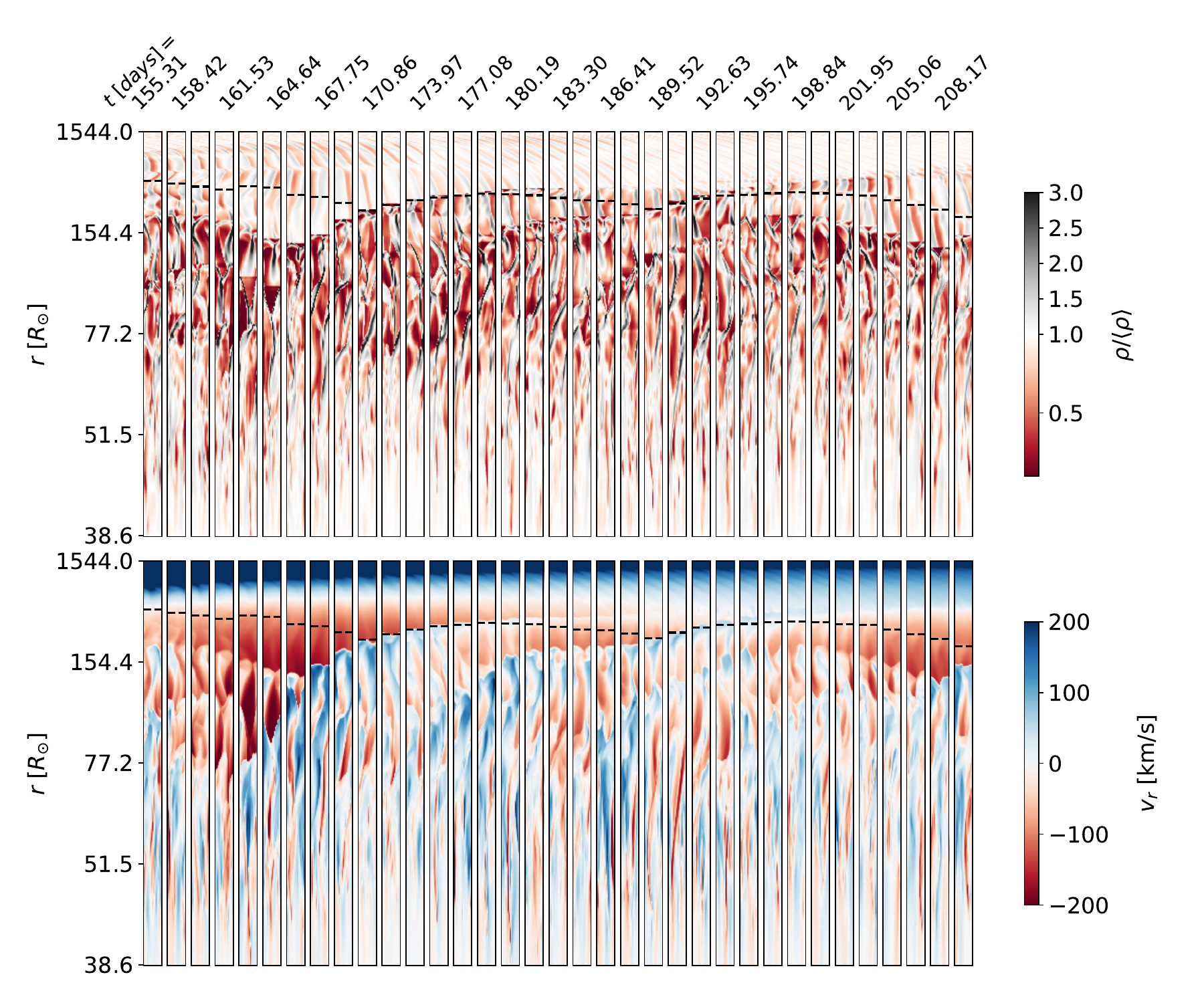}
  \caption{Time series for the coolplate simulation. The top panel shows relative density, the bottom panel shows the radial velocity. The radial direction is displayed using the $x \equiv 1 - R_c/r$ coordinate, which enhances the visibility of the inner regions. This is a clear example of the highly turbulent nature of the simulated atmospheres. The black dashed line on each panel denotes the location of the photosphere. 
  }
     \label{Fig:2D-slices_380}
\end{figure*}

\subsection{2D turbulent model with line-driven wind (`coolplate')} \label{Sect:Sub-Edd_sims}

Figure \ref{Fig:2D-slices_380} shows a (short) time series of 2D snapshots of the relative density (top panel) and the radial velocity (bottom panel) of the coolplate simulation, taken after relaxation from the initial conditions. 
The radial coordinate $r$, depicted on the y-axis, is plotted using a rescaled $x \equiv 1 - R_c/r$, which stretches the inner regions while compressing the contrary, thus increasing visibility in the parts of the simulation which show the greatest variability. The relative density is defined as $\rho/\langle \rho \rangle$, with $\langle \, \cdot \, \rangle$ denoting lateral averaging for every individual snapshot. This produces a clear contrast between the over- and underdense regions with respect to the average radial density profile within the simulation domain, showing a highly turbulent medium throughout the full simulation, both in the optically thick deeper envelope, and in the optically thin line-driven wind. 

Since this time series covers almost three cycles of the semi-periodic behaviour of the simulation, there are distinct differences between snapshots.
A cycle starts when gas is propelled upwards from the iron opacity bump. In this region, advection transport is quite efficient (up to 90\% of the energy is transported by advection in the lower parts of the simulation). This is contrary to earlier models of O- and WR-stars \citep{nico_2022b,Debnath24,nico_2025}, which showed inefficient enthalpy transport in this region (with efficiencies ranging from 1\% for dwarfs to 30\% for supergiants).
We can qualitatively understand this difference through a simple estimate of the relative importance between radiative and advective (`convective') energy transport \citep{grafener_2012, jiang_2015, schultz_2023, jiang_2023, Debnath24}. Specifically, if the radiative diffusion timescale $t_{\rm diff}$ is shorter than the relevant dynamical timescale $t_{\rm d,a}$, gas particles quickly adjust to their surroundings such that energy transport by means of advection is no longer efficient. This can be directly illustrated by writing out the radiative diffusion timescale as $t_{\rm diff} \sim (\tau/c) H_{\rm p}$, for optical depth $\tau$ and pressure scale height $H_{\rm p}$, such that $t_{\rm diff}/t_{\rm d,a} \sim (\varv/c) \tau$, where $\varv$ is a typical turbulent velocity in the region. Using the standard scaling $\tau \sim T^4/T_{\rm eff}^4$, where an `effective' temperature is defined by $\sigma T_{\rm eff}^4 \equiv L_\ast/(4 \pi r^2)$, it becomes immediately clear how this will be much more efficient for these simulations as compared to our previous ones that had significantly higher $L_\ast/r^2$ ratios.

A large portion of the material that is accelerated upwards from the iron opacity bump, falls back before line-driving can kick in to further accelerate it up to escape speed.
As the gas falls back, it deposits the (mechanical) energy it carried back into the region of the iron opacity bump while also impeding the outward transport of advective energy. 
This triggers another `failed wind outburst' and the start of new cycle. 
In atmospheric regions above this, efficient line-driving ensures that the outer parts of the simulation remain outflowing for the full duration of the simulation, visible as the dark blue stripe at the top of Fig. \ref{Fig:2D-slices_380}. The velocity of the material at the outer boundary is around 250 -- 350~km/s, which is above the local escape speed of around 150~km/s. This results in a consistent line-driven wind with a mass-loss rate $2 - 3 \times 10^{-5}~M_\odot/{\rm yr}$.

\begin{figure*}
\centering
\includegraphics[width=\textwidth]{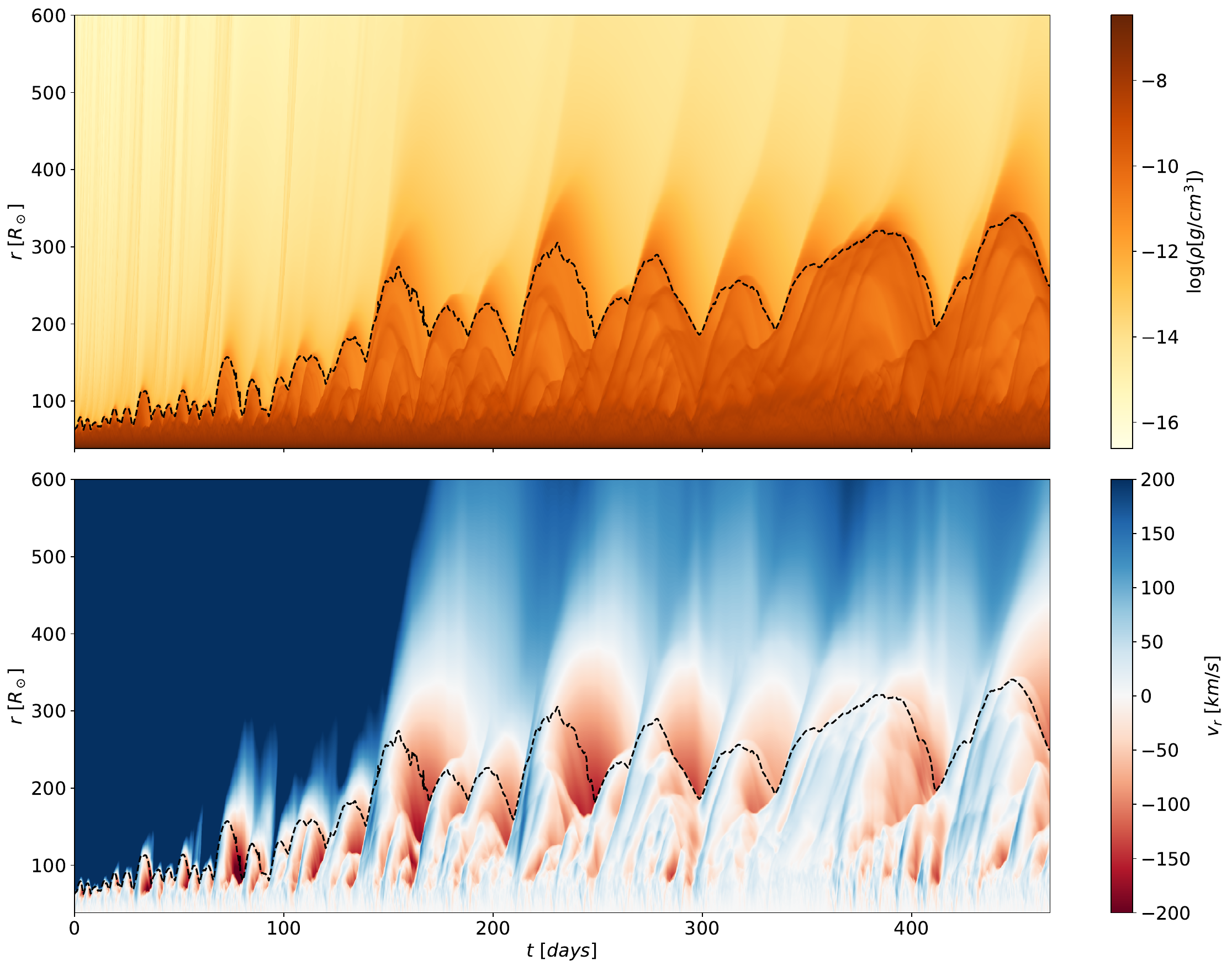}

  \caption{Space-time plots of the average density (top panel) and radial velocity (bottom panel) of the coolplate simulation, zoomed in to the lower part of the domain to increase visibility. The location of the photosphere is indicated by the black dashed line. 
   }
     \label{Fig:ST_vel_380}
\end{figure*}

By taking the lateral average of a quantity (i.e. producing a radial profile) and stacking every snapshot next to each other we produce a space-time plot that has a far finer time-resolution than Fig. \ref{Fig:2D-slices_380}. This sort of figure can then depict the full time-domain of the simulations, making it easier to spot general trends in the simulated data (at the cost of losing the detailed 2D view). An example of such a plot is shown in Fig. \ref{Fig:ST_vel_380}, which depicts the (absolute) density (top panel) and the radial velocity behaviour (bottom panel) of the coolplate simulation. Contrary to Fig. \ref{Fig:2D-slices_380}, the radial axis of Fig. \ref{Fig:ST_vel_380} is not transformed using the rescaled radial coordinate $x$, to depict the behaviour in a way that makes it more intuitive to interpret. The semi-periodic behaviour is very clear. After the relaxation of the initial conditions, large scale oscillatory behaviour develops, with `failed winds' reaching radii of 200 -- 300~$R_\odot$, before falling back down.
The upper parts of the domain remain optically thin for the entire duration of the simulation, so while some of the material that is thrown up is transported out in a line-driven wind, most of it falls down again. Since the `failed winds' consist of optically thick material, the photosphere, denoted with the black dashed line in Fig. \ref{Fig:ST_vel_380} and defined as the surface where the optical depth $\tau= 2/3$, also varies together with these `failed winds'. 
This creates an interplay between the deep envelope layers and the line-driven wind on top of it that is quite reminiscent of our earlier O-star simulations \citep{Debnath24, Delbroek2025, nico_2025}, except for the much larger semi-regular variation of the optical photosphere here.

\subsection{2D quasi-steady super-Eddington outflow model (`2D hotplate')} \label{Sect:2D_Super-Edd_sims}

\begin{figure*}
\centering
\includegraphics[width=\textwidth]{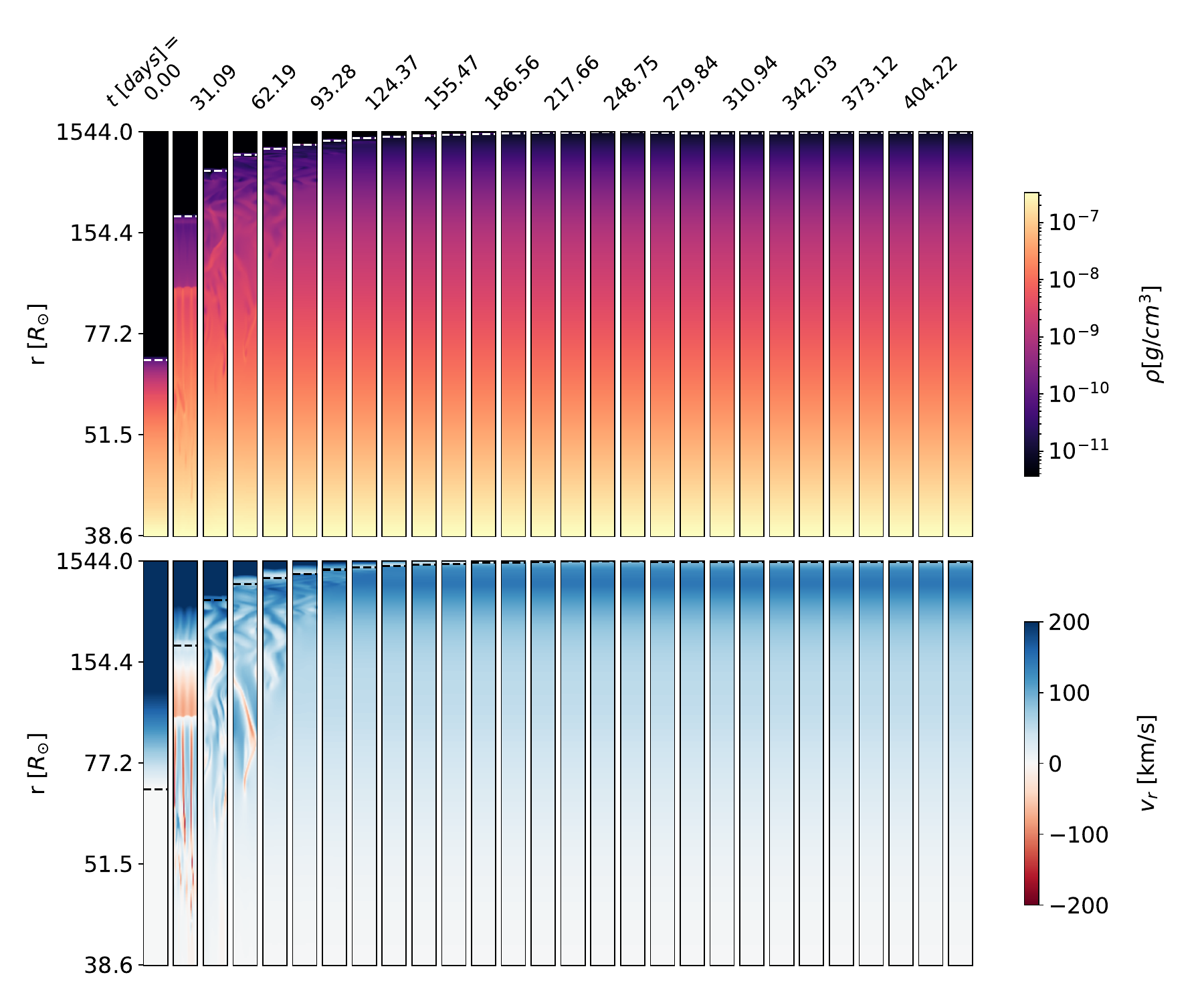}
  \caption{Same as Fig. \ref{Fig:2D-slices_380}, but for the hotplate simulation, and using density instead of relative density in the top panel. No turbulence is present after the initial relaxation.
   }
     \label{Fig:2D-slices_440}
\end{figure*}

Figure \ref{Fig:2D-slices_440} shows the same plot as Fig. \ref{Fig:2D-slices_380} for the hotplate simulation, except that the top panel shows the absolute density $\rho$, not the relative density. It is clear from the figure that this simulation occupies a very different regime than the coolplate one. Similar to the coolplate simulation, material is thrown upwards.
However, after the initial relaxation period, this simulation does not show the pronounced turbulence that was visible there, instead displaying a smooth, effectively quasi-stationary outflow that never reaches escape velocity, initiated from deep within the envelope.
This outflow transports high-density material upwards and has enough momentum and energy to carry this to the outer boundary of the simulation without it falling down, filling up the entire domain with optically thick material so that the photosphere is no longer resolved within the simulation domain. 

It is useful to interpret these simulations by considering a 1D, fully radial stationary state. 
We can then write a Bernoulli-type equation for conservation of total energy \citep{Cassinelli73, owocki_2017},
\begin{equation} \label{eq:1DenergyconsFtot}
  \rho \varv_r \Bigg( \frac{5}{2} \frac{P_g}{\rho} + \frac{E}{\rho} + \frac{P}{\rho} + \frac{\varv_r^2}{2} - \frac{GM}{r} \Bigg) + F_{\rm cmf} = F_{\rm tot},
\end{equation}
with $P_g$ the gas pressure, $\rho$ the density,  
$\varv_r$ the radial velocity, $F_{\rm cmf}$ the CMF flux and $F_{\rm tot}$ the total flux. Here, the enthalpic flux is given by

\begin{equation}
  F_{\rm enth} = \rho \varv_r (h_g + h_r) = \rho \varv_r \bigg( \frac{5}{2} \frac{P_g}{\rho} + \frac{E}{\rho} +  \frac{P}{\rho} \bigg),
\end{equation}

with $h_g$ and $h_r$ the gas and radiation enthalpy, respectively, assuming $\gamma_g =5/3$ for the former. These equations can be readily converted from fluxes to luminosities by multiplying by $4\pi r^2$, whereby the distinction between observer's frame and CMF radiative luminosity, $L_\ast = L_{\rm cmf} + \dot{M} h_r$, also becomes clear. In a stationary radial flow without energy source terms the total luminosity $L_{\rm tot} = 4 \pi r^2 F_{\rm tot}$ is constant. In addition to enthalpy and radiation, we also readily identify kinetic and gravitational energy fluxes through the terms $\varv_r^2/2$ and $GM/r$, respectively, from Eqn. \ref{eq:1DenergyconsFtot}.

Even though this model, with its higher total energy, (significantly) breaches the classical Eddington limit at the outer boundary, the lower parts of the simulation remain sub-Eddington as radiative enthalpy is responsible for the transport of most of the energy in these lower parts. However, the efficiency of enthalpy in carrying energy drops steeply when moving upwards in the atmosphere, causing the CMF flux to increase again in order to preserve the total energy, thus dominating the energy transport and raising the local Eddington factor. 
This then results in super-Eddington luminosities and drastic envelope inflation, driving a super-Eddington flow from within the optically thick envelope \citep[see also][]{Quataert2016, owocki_2017}. 

\begin{figure*}
\centering
\includegraphics[width=\textwidth]{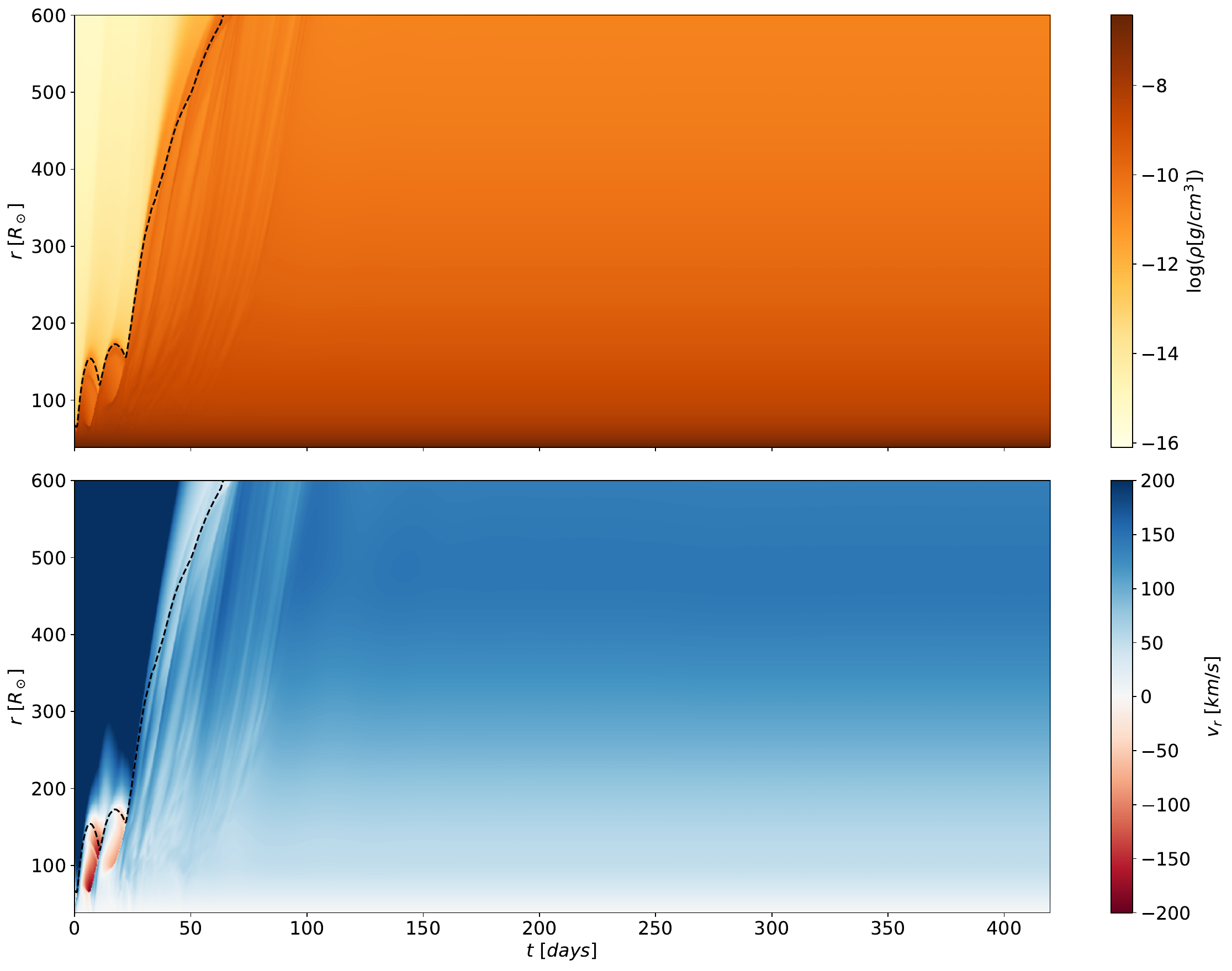}
  \caption{Same as Fig. \ref{Fig:ST_vel_380}, but for the hotplate simulation.
   }
     \label{Fig:ST_vel_440}
\end{figure*}

Figure \ref{Fig:ST_vel_440} shows the space-time plot for the density (top panel) and the radial velocity (bottom panel) of the hotplate simulation. Like in Fig. \ref{Fig:2D-slices_440}, this figure is at first dominated by the `outburst' triggering right after initialisation. The plot shows how the velocity does not reach escape velocity in this simulation, which is around 150~km/s at the outer boundary, as mentioned above. 
That means that while the high-density material is coasting out of the simulation domain at a rate of $\sim 10^{-1}~M_\odot/{\rm yr}$, this is all still gravitationally bound and thus does not constitute an actual mass-loss rate. To resolve the photosphere, capture the wind and get out a mass-loss rate, one needs to extend the simulation domain by a very large factor, which becomes computationally prohibitive in a multi-dimensional set-up. However, due to the quasi-stationary nature of these super-Eddington simulations (see Fig. \ref{Fig:2D-slices_440}, and also \citealt{owocki_2017}), this can be done here to a reasonable approximation in 1D. 

\begin{figure}
\centering
\includegraphics[width=\hsize]{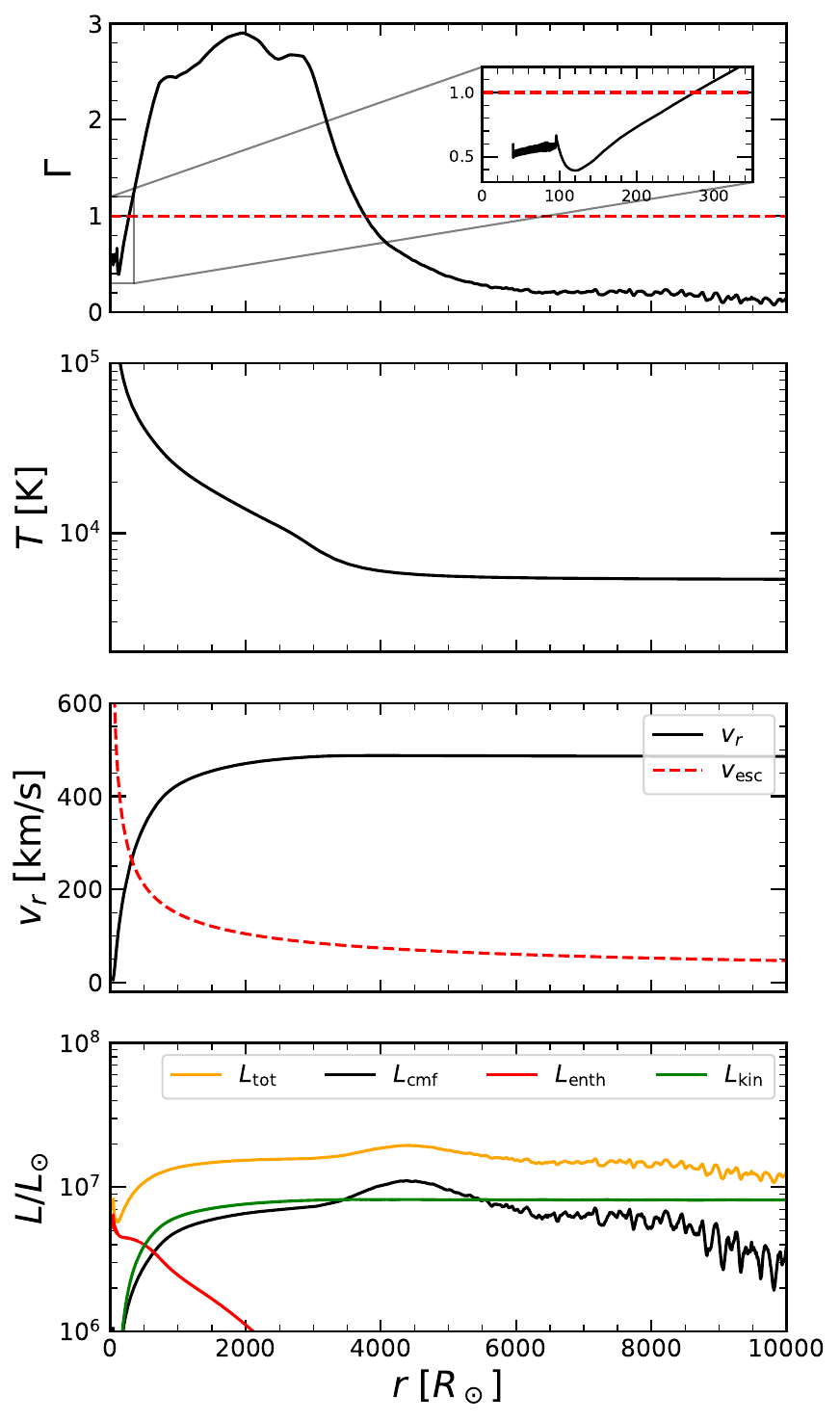}
  \caption{Time-averaged radial profiles of the 1D hotplate simulation. Top panel shows the Eddington ratio, second panel shows the temperature, third panel shows the radial velocity and the local escape velocity, and the bottom panel shows the total luminosity, CMF luminosity, radiative enthalpy transport and kinetic energy transport.
  }
     \label{Fig:1D-profiles_440}
\end{figure}

\subsection{1D quasi-steady super-Eddington outflow model (`1D hotplate')} \label{Sect:1D_Super-Edd_sims}

Some radial profiles of such a 1D simulation are shown in Fig. \ref{Fig:1D-profiles_440}. 
This figure shows the radial, time-averaged profiles (up to 10~000~$R_\odot$) of the Eddington ratio, temperature, radial velocity and luminosity of the 1D hotplate simulation, i.e. the 1D simulation with a lower boundary temperature of 440~kK. We can draw several conclusions from these profiles. 

Firstly, even though the CMF luminosity is extremely high throughout the simulation (approximately $8 \times 10^6~L_\odot$), the Eddington ratio $\Gamma$ is not above unity over the entire domain. Close to the lower boundary, the simulation starts sub-Eddington, only breaching the Eddington limit after around 280~$R_\odot$ (see inset in Fig. \ref{Fig:1D-profiles_440}). 
Then, roughly 4000~$R_\odot$ from the centre of the star, the Eddington ratio drops below one again, and stays low from that point out. 

Secondly, similar to the 2D models, the behaviour in the inner parts of the simulation is modulated by the competition between radiative enthalpy and CMF flux in transporting energy, where the former is higher in the lower parts whereas the latter completely dominates in the upper (see Fig. \ref{Fig:1D-profiles_440}).  
We can understand this transition by inspecting the ratio between the convective flux provided by radiative enthalpy and the diffusive CMF flux, which in the radiation-dominated regime scales as $F_{\rm enth}/F_{\rm cmf} \sim (\varv_r/c) (T/T_{\rm eff})^4 \sim (\varv_r/c) \tau$; indeed, this is the same basic scaling as discussed in the coolplate simulation when estimating the `efficiency' of convective energy transport. This gives a transition optical depth $\tau_t \approx c/\varv_r$ below which radiative enthalpy will not be efficient for transporting energy (see also discussion in \citealt{owocki_2017}, and in the context of convection where this is sometimes called the `critical' optical depth, see e.g. \citealt{schultz_2023, jiang_2023}). For the simulation visualised in Fig. \ref{Fig:1D-profiles_440} we have $\varv_\infty \approx 500$~km/s and may thus estimate $\tau_t \approx c/\varv_\infty \approx 600$. Inspection of the average optical depth in this simulation indeed confirms that the position at which $L_{\rm cmf} \approx L_{\rm enth}$ corresponds to $\tau$ of a couple of hundreds, supporting this order-of-magnitude scaling relation.  

Thirdly, we see that the outflowing wind indeed starts deep within the envelope of the star. In fact, the wind reaches a terminal velocity of approximately 500~km/s already after 2000~$R_\odot$, which is still within very optically thick layers. 
Due to the inflated nature of these simulated stars, the temperature goes down to below the hydrogen ionisation temperature before the atmosphere becomes optically thin. That means that the material does not experience any further acceleration, as when hydrogen recombines, the opacity drops to effectively close to zero. Instead, matter just continues moving outwards at this velocity. 
In contrast to the 2D hotplate simulation, material reaches velocities higher than the escape velocity (which is only around 75~km/s at the optical photosphere here).
The steep drop in opacity when hydrogen recombines at $T \sim 5000-5500$~K always determines the position of the photosphere in these simulations; that is, we will essentially always have $T_{\rm eff} \sim 5000-5500$~K in this regime.
As line-driving at such low temperatures is very inefficient, there is no further acceleration of the wind in the optically thin outer parts. 
While an additional boost from dust-driving  \citep{Hofner18} could potentially play a role even farther away from the star, when the temperature has dropped even more, this is not considered here. 

Lastly, the kinetic luminosity $L_{\rm kin} = \dot{M}\varv_r^2/2$ and the observer's frame luminosity $L_{\ast} = L_{\rm cmf} + L_{\rm enth} \approx L_{\rm cmf}$ are of similar magnitude around the optical photosphere, $L_\ast \approx L_{\rm kin} \approx 8 \times 10^6$~L$_\odot$, with L$_{\ast}$ varying less than a factor two in the relevant layers around the photosphere (Fig. \ref{Fig:1D-profiles_440}). Thus, even though $L_{\ast}$ is super-Eddington the outflow is somewhat `photon tired' \citep{owocki_gayley_97}. 

\begin{table}
\centering
\caption{Overview of the different simulations with resolved photospheres.} \label{Table:overview}
\begin{tabular}{cccc}
\hline \hline
$T_c$ [kK] & $L_\ast / L_{\rm Edd}$ & Dimension & Notes       \\ \hline
360            & 0.56                                                     & 2D        &             \\
380            & 0.62                                                     & 2D        & `coolplate' \\
400            & 0.82                                                     & 2D        &             \\
420            & 0.98                                                     & 2D        &             \\
440            & 18.41                                                    & 1D        & `hotplate'  \\
450            & 21.86                                                    & 1D        &             \\
460            & 33.49                                                    & 1D        &             \\ \hline
\end{tabular}
\end{table}

\subsection{Overview of resulting parameters for the model grid} \label{Sect:Overview_grid}

\begin{figure}
\centering
\includegraphics[width=\hsize]{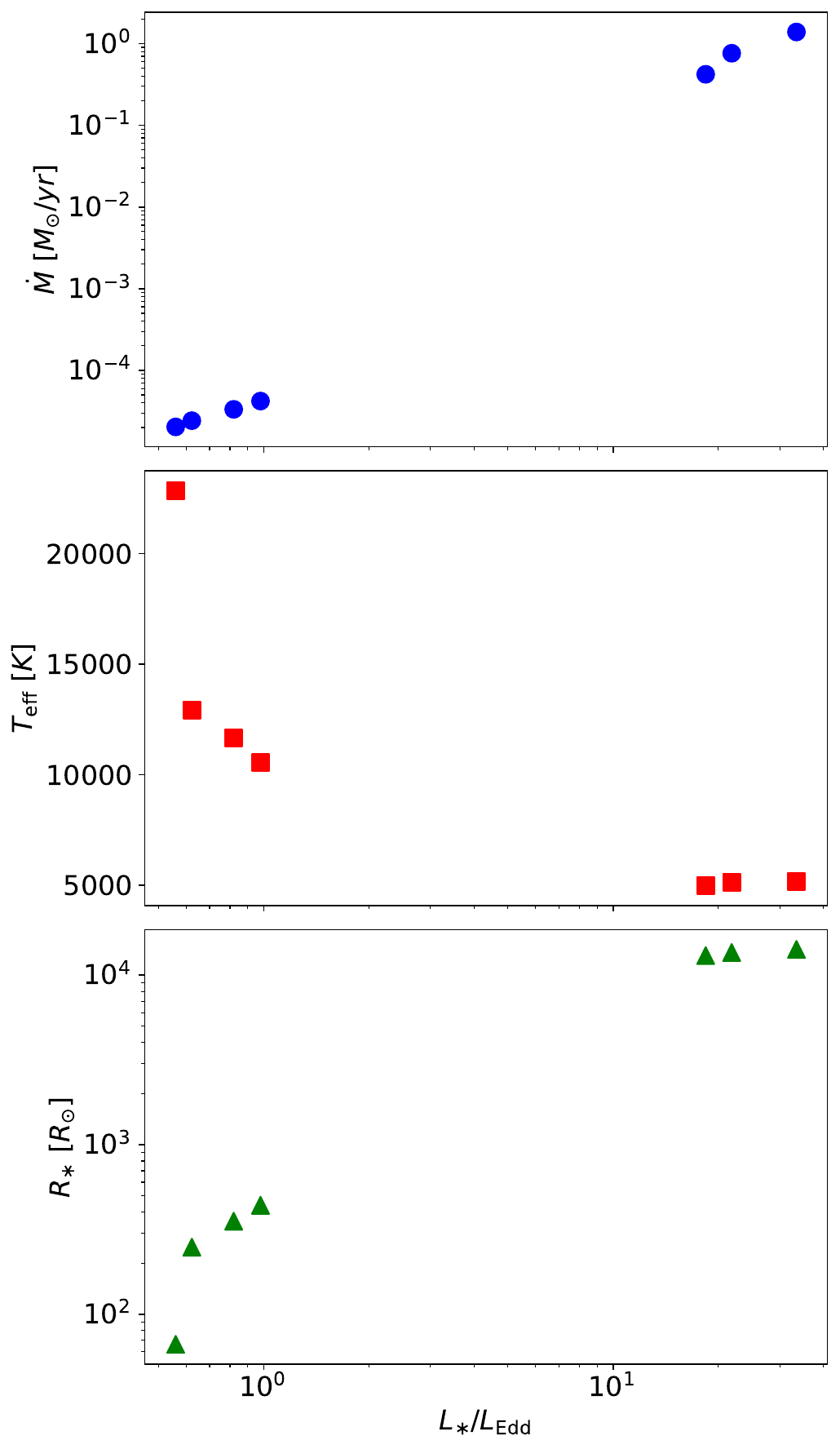}
  \caption{Change of average $\dot{M}$, $T_{\rm eff}$ and $R_*$ with respect to the classical Eddington factor. A clear transition can be seen from the sub-Eddington to the super-Eddington regimes.}
     \label{Fig:series_progression_2D_only}
\end{figure}

Figure \ref{Fig:series_progression_2D_only} shows an overview of the seven simulations with a resolved photosphere, of which the three most energetic are 1D. This overview is supplemented by Table \ref{Table:overview} and Fig. \ref{Fig:HRD}.
The models are plotted against the ratio between observed luminosity $L_\ast$ at the photosphere and the classical Eddington luminosity $L_{\rm Edd}$. 
Here, it should be noted that the difference between $L_\ast$ and $L_{\rm cmf}$ at the photosphere is on average only 2 -- 4~\%, and never exceeds 10~\%.
The panels show, from top to bottom, the averaged mass-loss rate $\dot{M}$, effective temperature, and photospheric radius.

This series shows a clear transition from sub-Eddington to super-Eddington photospheric conditions. The sub-Eddington simulations have turbulent envelopes with a line-driven wind on top, with mass-loss rates between 2 and $5 \times 10^{-5}~M_\odot/{\rm yr}$.
Their effective temperatures go from around 22.5~kK, for the least extreme star, to slightly above 10~kK, dropping off steeply. The photosphere expands, from 66~$R_{\odot}$ to nearly 450~$R_{\odot}$. We note that these quoted mass-loss rates only count material that has reached the local escape speed. Closer examination of the wind launching region around the optical photosphere in our models shows that the mass-loss rate there is extremely variable and often much higher, reaching instantaneous values of $\dot{M} \sim 10^{-2} - 10^{-3} \ \rm M_\odot/year$ and average rates of $\dot{M} \sim 10^{-3} - 10^{-4} \ \rm M_\odot/year$. However, since a significant portion of this never reaches escape speed we do not count it in our quoted mass-loss rates. However, this likely has important consequences when interpreting empirically inferred mass-loss rates from LBVs, as these are always derived from models assuming a stationary pure outflow (e.g. \citealt{Najarro12}).

For simulations with super-Eddington photospheric luminosities, this all drastically changes. The turbulence in the deep atmosphere is no longer present in these models (not in the 2D ones either, see Fig. \ref{Fig:2D-slices_440}), and the mass loss becomes on the order of $10^{-1} - 1~M_\odot/{\rm yr}$.
The effective temperature for all these models is around 5000~K (see above), and their atmospheres inflate to $\sim 10~000~R_{\odot}$. 

It is notable that the classical Eddington factor in this series does not just increase as $\sim \Delta T_{\rm c}^3$. 
Instead, there is a very sudden and explosive switch, as none of the super-Eddington models have photospheric classical Eddington ratios less than ten. 
This transition, which comes out naturally when increasing the total stellar energy in these RHD simulations, essentially progresses from luminous blue hypergiant-like models to models that show giant eruption/supernova impostor-like behaviour. This means that these different observational phenomena may not be distinct categories, but stem from similar objects receiving less or more energy in this critical stage of their evolution.

\section{Discussion and conclusion} \label{Sect:Discussion}
The central result from the set of simulations presented here regards the drastic transition in overall atmosphere and wind properties as stellar energy is gradually increased. Specifically, the increase in total stellar energy results in a change in the structure of the atmosphere and the key characteristics of the resulting emergent parameters, going from a highly turbulent, moderately hot photosphere with a persistent line-driven wind on top, to a quasi-steady super-Eddington wind with a recombined cool photosphere. The distinction between these two categories is caused by the total luminosity in the models. When the total luminosity is too low for the material to be propelled upwards enough to reach local escape velocity, the material falls back. In this case, the atmosphere experiences deep-seated, semi-periodic `failed winds' that influence the photospheric radius, but is not the key driver of mass loss, as that is set by efficient line-driving from the turbulent photosphere. Wind velocities in this regime are typically $\varv_{\infty} \approx 200-300$ km/s and average mass-loss rates $\dot{M} \approx 2-5 \times 10^{-5} \rm \, M_\odot / year$. Such values are in reasonable agreement with observations of quiescent S Doradus-type LBVs such as P Cygni \citep{Najarro12}. We note that the instantaneous mass-loss rate launched from just above the photosphere in our models approaching (but not breaching) the Eddington limit often is substantially higher than our quoted average values, but how a significant portion of the launched gas never reaches the local escape speed and thus falls back upon the star. This likely has important consequences for interpreting observations and empirically inferred mass-loss rates of LBVs, as such studies are always based on models assuming a stationary pure outflow.

On the other hand, when the luminosity becomes high enough, radiative enthalpy initiates a flow from the deep envelope that is able to reach escape speed. 
In this case, the atmosphere inflates massively and a successful super-Eddington wind is launched, with the material coasting out at a relatively low velocity from the optical photosphere determined by the opacity drop when hydrogen recombines.
Such slow-moving, super-Eddington winds with cool photospheres are also observed, e.g. in $\eta$ Carinae's 19th century `great eruption'. Using light echos, \citet{rest_2012} inferred ejecta velocities of a couple of hundreds of km/s and a cool photosphere on the order of $T_{\rm eff} \approx 5000$ K from the observed spectra. Moreover, during this giant eruption $\eta$ Carinae was observed to be highly super-Eddington at the photosphere and with an extreme mass-loss rate of $\dot{M} \sim \rm 1~M_\odot/year$ \citep{Davidson12}, resulting in values of wind kinetic power and observed radiative luminosity that are comparable to one another. Although our models here were not tuned to an $\eta$ Carinae-like star specifically, they do agree qualitatively with these observations of high mass loss, slow moving material coming from a cool object with an observed super-Eddington luminosity.

As mentioned in previous sections, in this paper we have focused on characterising the atmosphere and wind properties resulting from an assumed energy increase in the stellar envelope. For $\eta$ Carinae, an energy increase has been suggested to be triggered by a merger event \citep{Hirai21} and repeated binary interactions have also been suggested as a potential origin for the quasi-periodic outbursts observed for some supernova impostors \citep{Aghakhanloo25}. It is presently not clear under which conditions energy increase might occur from the internal evolution of the massive star itself and when such external triggers are needed.

\subsection{Photospheric oscillatory behaviour}

The photosphere oscillates in all simulations presented here. The amplitudes vary from around 20~$R_\odot$ for the coolest coolplate simulation to around 10~000~$R_\odot$ for the hottest hotplate simulation. 
In general, oscillations are expected for almost all stars (see e.g. \citealt{Asteroseismology_book_Conny}), and specifically in a fairly similar parameter regime as the (sub-Eddington) regime investigated here. Earlier work by \citet{jiang_2018,grassitelli_2021, goldberg_2025} showed (semi-periodic) photospheric variability on the order of at most a couple of hundred $R_\odot$. This is on the same order as found here in the 2D simulations that stay below the Eddington luminosity (see Fig. \ref{Fig:ST_vel_380}). 

On the other hand, our 1D super-Eddington simulations sometimes experience giant oscillations with much larger amplitudes,  where the photospheric radius can quickly (in the course of days) increase, and then again decrease, by a factor of approximately four. \citet{Eva25} report similar behaviour in 1D dynamical interior simulations of luminous pre-supernova red supergiants, finding extreme oscillations (factor three) of the photospheric radius. Over the course of these giant radial oscillations, we still only find moderate changes in photospheric effective temperature, not greater than around 1000~K, reflecting again the fact how the photosphere in these simulations is determined solely by hitting the opacity-wall when hydrogen ionises (see discussion above).

The oscillations of these two classes of simulations are caused by different physical processes. The origin of the oscillations in the sub-Eddington models was already discussed in Sect. \ref{Sect:Sub-Edd_sims}; in short, gas parcels are thrown upwards from deep layers but stagnate as they do not have enough momentum to escape the atmosphere. When the parcels start to fall back into the envelope again, they collide with upflows causing heating of the deeper parts, triggering a new cycle. The giant oscillations in the 1D super-Eddington simulations, in contrast, are triggered by the reionisation of hydrogen in a partial ionisation zone. 
The upper parts of the outflow are heated, causing the hydrogen ionisation front to move outwards. This increases the opacity locally by over an order of magnitude, quickly moving the photosphere (i.e. where the optical depth $\tau = 2/3$) outwards. This increase in opacity (and thus radiative force) does not lead to re-acceleration of the outflow since $\Gamma$ remains well below one for the entire range in radius where this occurs.

It is important to note, however, that these giant oscillations do not seem to be universally present in super-Eddington models.
We constructed several simulations with different lower boundary conditions than the simulations presented in earlier sections. Specifically, we tested different combinations of lower boundary radius, temperature, and stellar mass. While the simulations differ in details, the general structure of the super-Eddington flow described above is still valid, with $\Gamma$ remaining below one at the lower boundary and as long as radiative enthalpy is efficient in transporting energy, with again photospheric $T_{\rm eff} \sim 5000$ K due to hydrogen recombination, and the radial velocity reaching its terminal value, which is higher than escape velocity, well below the photosphere. However, the giant oscillations do not always occur, and the time-averaged photospheres of such simulations lie generally between 5000 and 10~000~$R_\odot$.

\subsection{Lower boundary conditions}

The main simulations presented in this paper were carried out by varying the temperature at an otherwise fixed lower boundary. 
As demonstrated, the final states of the simulated stars are very sensitive to this variation, where relatively small variations can induce large changes in outcome. By varying only the total stellar energy and keeping the rest constant, the aim is to effectively mimic approximately how a star would behave when moving upwards in the HRD toward higher and higher $L/M$ ratios.
The effect of the dynamic atmosphere on the rest of the underlying stellar structure is here ignored (as is also done in other dynamical simulations, e.g. \citealt{jiang_2018,Eva25,goldberg_2025}). 
However, in this extreme parameter regime it is possible that feedback from the outer envelope, atmosphere, and wind upon the deeper core layers might lead to rather fast re-adjustment of also the stellar interior \citep{grassitelli_2021}. 
As such, future work in this domain would likely benefit from including a feedback mechanism that couples the behaviour of the stellar atmosphere to the rest of the star, in order to better assess the response of the underlying star to its eruptive surface behaviour. Such a feedback mechanism may bring further clues about under which conditions the eruptive behaviours observed for stars in this part of the HRD can be explained solely by internal evolution and feedback from its atmosphere and wind, and under which conditions extra energy-input (from e.g. binary interactions) is needed.

\subsection{General limitations of current simulations} \label{Sect:Discussion:Limitations}

As mentioned in Sect. \ref{Sect:Method}, we fixed the line-force parameters to the canonical values $\bar{Q} = 2000$, $Q_0 = 2000$ and $\alpha = 2/3$ \citep{gayley_1995}. This ensures that the line-driven wind remains outflowing at a reasonable rate, making the simulations relax more quickly from their initial conditions and reducing numerical difficulties and the computational cost in this challenging regime. More realistically, however, the line-force parameters should be computed for the local and time-varying wind conditions \citep{nico_2022b, Debnath24}. A simple application (using $Z = 0.02$) of our line-driven-wind iterative mass-loss estimator (\href{https://lime.ster.kuleuven.be/}{LIME}, \citealt{Sundqvist25}) yields for the hottest simulated star presented here ($T_{\rm eff} \approx 22.5$ kK) $\bar{Q} = 2000$, $Q_0 = 3400$ and $\alpha = 2/3$ at the wind critical point, which is not too far away from our assumed canonical values. However, the corresponding analytically predicted line-driven mass-loss rate is significantly lower than what we find from our 2D simulations. It is likely that this mismatch is because turbulent pressure yields an effective gravity in the wind-launching region of our highly inflated 2D simulations that is significantly lower than in  \href{https://lime.ster.kuleuven.be/}{LIME} (which only takes into account standard reduction of effective gravity due to electron scattering). This should be further investigated by future work targeting the interplay between a turbulent atmosphere and line-driven wind in the AB-hypergiant regime \citep{Bernini26}. Additionally, we are here also in a parameter regime where non-local thermal equilibrium corrections may be necessary to include in order to obtain more realistic line-force parameters (Appendix A in \citealt{Sundqvist25}). Work is under way in our group to include this in multi-dimensional simulations, and results will be presented in upcoming papers. 

We assume a fixed adiabatic index for the gas, $\gamma_g = 5/3$. While this likely is a very good assumption for stars where hydrogen remains ionised at the surface, it may potentially affect our super-Eddington models with recombined, cool surfaces. 
In the multi-dimensional simulations of yellow and red supergiants by \citet{goldberg_2022, goldberg_2025} they also assume constant $\gamma_g = 5/3$; however, in the 1D dynamical pre-supernova red supergiant models by \citet{Eva25} the authors find that ionisation energy in the partial hydrogen (and helium) recombination zones play a significant role in controlling the oscillating stellar surface. Comparable models for the high-luminosity regime investigated in this paper are lacking, however.  

We assume the FLD closure relation as outlined by \citet{nico_2022a} in all our RHD simulations. This means we only solve the radiation energy equation, and thus radiative momentum is technically not a conserved quantity within our framework. This may partly cause the variable radiative fluxes we observe also in our simulations with (quasi-)steady outflows (Fig. \ref{Fig:1D-profiles_440}). Work is currently under way by one of us (N. Moens) to replace the FLD approximation by a so-called `M1' closure \citep{Gonzalez2007} that ensures also radiative momentum conservation. While this will be a significant improvement as compared to our current simulation method, we do not expect it to affect significantly the overall conclusions of this paper.

\begin{acknowledgements}
The computational resources used for this work were provided by Vlaams Supercomputer Centrum (VSC) funded by the Research Foundation-Flanders (FWO) and the Flemish Government. The authors gratefully acknowledge support from the European Research Council (ERC) Horizon Europe under grant agreement number 101044048, of the Belgian Research Foundation Flanders (FWO) Odysseus program under grant number G0H9218N, of FWO grant G077822N, and of KU Leuven C1 grant BRAVE C16/23/009. The authors would like to thank all current members, and several former ones, of the KUL-EQUATION group for fruitful discussions, comments and suggestions. PS additionally wishes to express his sincere appreciation to Stan Owocki for many valuable insights and enlightening discussions. We made significant use of the following packages to analyse our data: \texttt{NumPy} \citep{harris_2020}, \texttt{SciPy} \citep{virtanen_2020}, \texttt{matplotlib} \citep{hunter_2007}, \texttt{Python amrvac\_reader} \citep{keppens_2020}.

\end{acknowledgements}

\bibliographystyle{aa}
\bibliography{references} 

\label{LastPage} 
\end{document}